\documentclass[review]{elsarticle}

\usepackage{lineno,hyperref}
\usepackage{amsmath}
\usepackage{amsfonts}
\usepackage{amssymb,amsthm,bm,multirow}
\usepackage{color}
\usepackage{subfigure}
\usepackage{soul}
\newcommand{\mb}{\mathbf}

\journal{Renewable Energy}

\begin{document}
	
	\begin{frontmatter}
		
		\title{A Case Study of Space-time Performance Comparison of Wind Turbines on a Wind Farm}
		
		\author{Yu~Ding\corref{cor1}}
		\ead{yuding@tamu.edu}
		\author{Nitesh~Kumar\corref{cor2}}
		\author{Abhinav~Prakash\corref{cor2}}
		\author{Adaiyibo~E.~Kio\corref{cor2}}
		\address{Department of Industrial \& Systems Engineering, Texas A\&M University, College Station, TX 77843, USA.}
		\cortext[cor1]{Corresponding author}
		
		\author{Xin~Liu\corref{cor3}}
		\ead{liuxin1@tianrun.cn}
		\author{Lei Liu\corref{cor2}}
		\author{Qingchang~Li\corref{cor2}}
		\address{Beijing TianRun New Energy Investment Corporation Ltd., F22-23,CSC Fortune International Center, No.5 An'ding Road, Chaoyang District, Beijing 100029, China}
		\cortext[cor3]{Co-Corresponding author}

		

		
		
		
		
	\begin{abstract}
			This paper presents an academia-industry joint case study, which was conducted to quantify and compare multi-year changes in power production performance of multiple turbines scattered over a mid-size wind farm. This analysis is referred to as a space-time performance comparison. One key aspect in power performance analysis is to have the wind and environmental inputs controlled for. This research employs, in a sequential fashion, two principal modeling components to exercise tight control of multiple input conditions---a covariate matching method, followed by a Gaussian process model-based functional comparison. The analysis method is applied to a wind farm that houses 66 turbines on a moderately complex terrain.  The power production and environmental data span nearly four years, during which period the turbines have gone through multiple technical upgrades. The space-time analysis presents a quantitative and global picture showing how turbines differ relative to each other as well as how each of them changes over time.
		\end{abstract}

		
	\end{frontmatter}
	

	\section{Introduction}\label{section1}
	Evaluation and comparison of the power production performance of wind turbines are important to many wind farm owners/operators, as power production capacity is invariably one of the key performance indices (KPI) and plays crucial roles in numerous decisions made routinely by the owners/operators. We want to clarify that (a) power production performance, referred to in this study, is pertinent to commercial production of wind turbines, rather than during the design or testing stages and (b) this performance evaluation treats an individual turbine as a holistic power production unit, with wind and other environmental conditions impacting a turbine as the inputs and power produced at that turbine as the output.  There are two primary aspects of the performance evaluation mission---to compare one turbine with another turbine on the same wind farm and to compare a turbine with itself over time, especially during the periods when certain technical upgrades have been undertaken. Addressing both the space and time aspects, we refer to the performance evaluation and comparison, reported in this paper, as a space-time performance analysis.
	
	Wind energy is a variable renewable energy source because wind and environmental conditions are changing all the time. For this reason, in the task of turbine performance comparison, it is critical to have the input conditions controlled for before comparing the power outputs.  The International Electrotechnical Commission (IEC) standard practice recognizes the intermittent and stochastic nature of wind and, therefore, recommends the use of the power curve for the purpose of evaluation~\cite{IEC12,IEC26-2}. The topic of turbine performance evaluation is covered in Chapter 6 in~\cite{Ding2019} as well as in~\cite{Niu2018}.  Specialized quantification methods, especially for vortex generator (VG) installation, are explained in Chapter 7 in~\cite{Ding2019}, or the originally published papers~\cite{Lee2015a, Shin2018}.
	
	Many of the existing turbine-level or turbine-specific performance evaluation methods, especially those based on the IEC standard practice, primarily control one principal input factor, which is the wind speed.  This is not to say that the IEC does not recognize the influence of other conditions. Rather, the point is that the conventional way of accounting for other factors is less effective.  On this point, Chapter 5 in~\cite{Ding2019} provides an elaboration backed by numerical evidence. Another shortcoming of the existing performance evaluation methods is the fact that although the power curve is a functional curve, or a functional response surface while considering multi-dimensional inputs, almost all existing methods reduce the functional curve to a scalar metric, be it the annual energy production (AEP)~\cite{IEC12}, or the power coefficient~\cite{Kjellin2011,Krogstad2012}, or a recently proposed productive efficiency measure~\cite{Hwangbo2017}.  We believe that it would be ideal to compare two functional curves directly without reducing them to a scalar metric because in comparing two functional curves, one can identify the regions of difference in power production, which can lead to valuable clues about performance changes and help figure out the root causes of these changes.
	
	We would like to address these technical issues for conducting the space-time performance analysis. To control for the multiple changes in wind and other environmental conditions, we adopt a covariate matching method~\cite[Chapter 7]{Ding2019}. Here the term \emph{covariate} refers to the wind and other environmental input conditions and is synonymous to the term \emph{input variable}. Through the process of matching, the input conditions for a turbine before and after a check point, or those for two turbines, become probabilistically comparable, as if these covariates were designed by experimenters to be drawn from the same probability distribution. As a result, the outcomes of two turbines of the same period or two periods of the same turbine are comparable under the matched covariate conditions.
	
	The covariate matching is the first main component of our proposed procedure.  Due to practicality constraints, covariate matching only controls the wind and other environmental conditions up to a prescribed discrepancy threshold, say, 25\%.  The second main component is to conduct a direct functional curve comparison of multi-dimensional power curves.  This second component is based on a recent methodology development in the area of functional data analysis, and specifically, the Gaussian process (GP)-based method presented in~\cite{Prakash2020}.
	
	One may question why the first component is still necessary, considering the availability of the functional curve comparison method in the second component.  The reason is that the method used in~\cite{Prakash2020}, although theoretically capable of comparing functional curves of any dimensions, needs, in practice, a large amount of data.  It also demands heavy computation if it is applied to high-dimensional data. It works, however, well enough for handling two or three input variables, e.g., wind speed and wind direction, or wind speed, temperature (or air density), and turbulence intensity.  The first component, on the other hand, can be conducted rather efficiently, even if there exist many input variables. Apparently, the inclusion of the first component is to have the varying inputs more or less controlled for, regardless of the inputs dimensions, before the second component is applied to the inputs of reduced dimension for functional comparison.
	
	The proposed method is first applied to two cases of known turbine upgrades.  The datasets used are publicly available. The results of our method are compared with two existing methods~\cite{Lee2015a, Shin2018} on upgrade quantification to establish credibility. We then apply the proposed procedure to data collected on 66 turbines, on a terrain of moderate complexity, over a four-year period.  The final comparison results are shown in a space-time illustration visualizing the quantitative and global picture of how turbines differ relative to each other as well as how each of them changes over time. All the analysis in the paper is done using \texttt{R} software~\citep{R2019} and the \texttt{DSWE} package~\citep{DSWE-package} available on \texttt{GitHub}.
	
	The rest of the paper unfolds as follows.  Section~\ref{section2} explains the application background and datasets to be used in this study. Section~\ref{section3} outlines the overall procedure for the proposed performance analysis and provides more details about the two important components in our procedure. Section~\ref{section4} presents the analysis results and discusses the practical implications.  Finally, Section~\ref{section5} summarizes this study.
	
	\section{Background and Data}\label{section2}
\begin{figure}[ht]
		\centering \includegraphics[width=\textwidth]{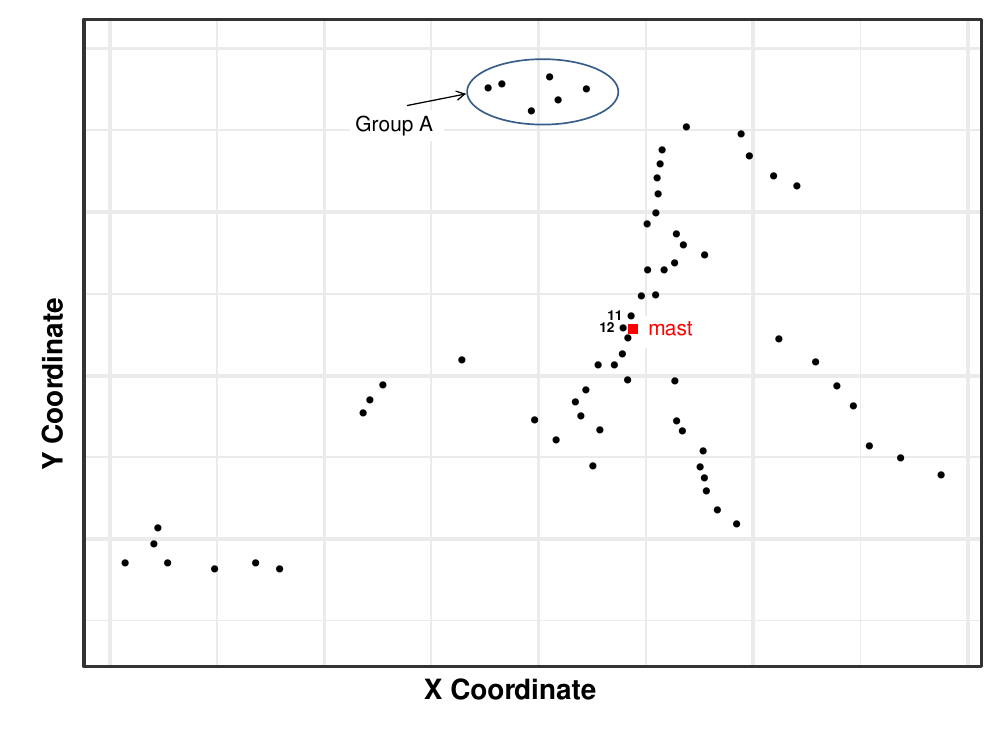}
		\caption{Layout of the turbines on the wind farm.}\label{figure1}
	\end{figure}

	Figure~\ref{figure1} presents the layout of the 66 inland turbines on the wind farm.  All turbines are of the same model, which belongs to a 1.5 MW turbine class.  The wind farm terrain is of moderate complexity. The elevation of the wind farm is not even.  For instance, the turbines in Group A are on a hill top, higher than the other turbines.  There is a single meteorological (met) mast on this wind farm.  Its location is near Turbine \#12. The wind farm layout has been transformed to protect the identity of the wind farm. But the relative positions between turbines are maintained to reflect the reality. To give a sense of the physical distance, the distance between \#12 and met mast is 80 meters and that between \#12 and \#11 is about 300 meters.
	
	The datasets collected on this wind farm run from August, 2014 through May, 2019, spanning a little bit over five years.  There are lots of missing data in the beginning phase of the period---between November, 2014 and February, 2015.  There are still missing data in the rest of the years, but less frequently so.  The data we end up using is effectively from July 2015 onwards for a period of about four years.
	
	The instant active power of a turbine is denoted by $y$.  The wind and other environmental covariates are denoted by the vector $\boldsymbol x$.  The datasets recorded wind speed, $W$, wind direction, $D$, and ambient temperature, $T$, but did not record air pressure or humidity. Because there is no air pressure measurement, temperature is used in place of air density. The measurements were taken on the nacelle of individual turbines using appropriate instruments, such as anemometer for measuring wind speed, wind vane for wind direction and thermometer for ambient temperature.
	
	The original datasets recorded the data at a frequency of seven to eight data points per second.  This high-frequency data is used to calculate turbulence intensity, denoted by $TI$, the standard deviation in wind direction, denoted by $sdD$, as well as to generate the 10-min average wind speed, direction and temperature.  Ultimately, $\boldsymbol x$ has five elements, i.e.,
	
	\vspace{-12 pt}
	\[\boldsymbol x=(W,\,\, T, \,\, D, \,\, TI, \,\, sdD)^\top.\]
	
\begin{figure}[t]
		\centering \includegraphics[width=.95\textwidth]{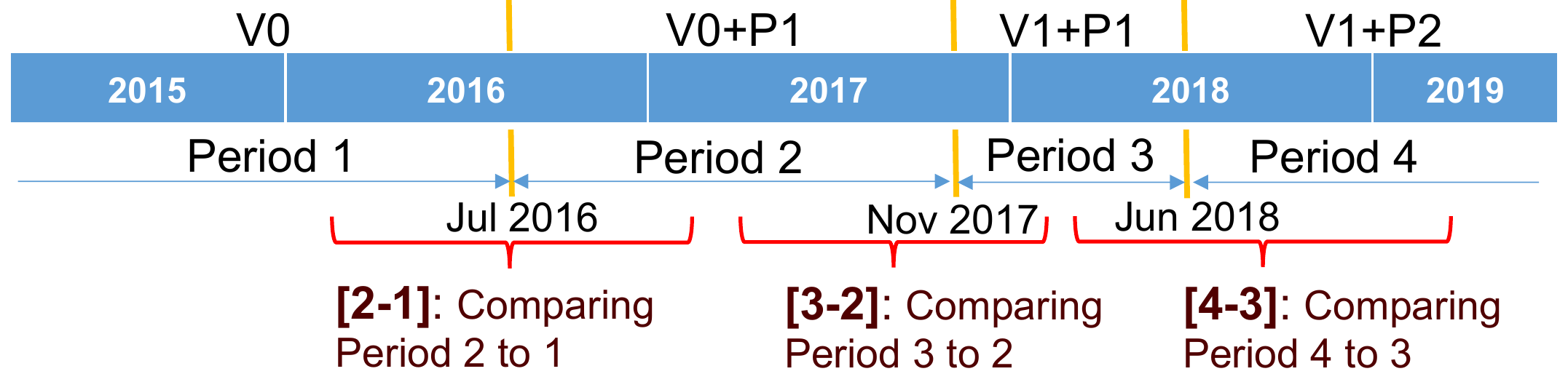}
		\caption{The timeline of the technical upgrades.}\label{figure2}
	\end{figure}

	The wind farm went through three technical upgrades during the four years, which partition the whole time domain into four periods.  Figure~\ref{figure2} illustrates the timeline of those upgrades. The nature of the upgrades is kept confidential, but whenever an upgrade was applied, it was carried out uniformly on all turbines. The three upgrades are coded as P1, V1, and P2, respectively. The times at which these upgrades were done is indicated in Figure~\ref{figure2}. In the figure, the code names in the top row indicate the technical setting of the turbines in the respective period. When we compare the turbine performance for Period 2 to 1 (denoted as ``[2-1]" in the Figure~\ref{figure2}), we are essentially comparing the technical settings ``V0+P1" and ``V0".  Other notations and dates can be likewise interpreted. The average number of data points per turbine for each period is reported in Table~\ref{table1}. The third period has fewer data points because it is shorter than the other three periods---it is about half a year, whereas the other periods are about a year.
	
	\begin{table} [b]
		\centering
		\caption{Number of data points per turbine corresponding to each period.} \label{table1}
		\resizebox{\textwidth}{!}{
			\begin{tabular}{lcccc}
				\hline  &Jul 2015--Jul 2016  & Aug 2016--Oct 2017 & Dec 2017--May 2018 &Jul 2018--May 2019 \\
				\hline \hline
				Technical setting & V0 & V0+P1& V1+P1& V1+P2\\
				Data amount & 47,712 & 52,373 & 22,800 & 37,658\\
				\hline
		\end{tabular}}
	\end{table}
	
	\section{The Method}\label{section3}
	This section describes the key components in the devised procedure.  The procedure is supposed to work on two sets of turbine data at a time. In terms of time-wise comparison, the data of two periods of the same turbine before and after a technical action are used to quantify how much a turbine changes due to that technical action.  In terms of space-wise comparison, the data of two turbines of the same period are used to quantify how much these two turbines differ from one another.  For each pair-wise comparison, three steps are undertaken sequentially:
	\begin{enumerate}
		\item Select the best covariate subset that contributes the most towards power curve modeling and rank the importance of the selected covariates.
		\item Conduct covariate matching to filter the data, so that the chosen subsets are statistically comparable.
		\item Use the data chosen in Step 2 to establish GP-based multi-dimensional power curve models, conduct the functional curve comparison, and quantify the difference.
	\end{enumerate}
	The three steps are explained in Sections~\ref{section3.1}--~\ref{section3.3}, respectively, whereas Section~\ref{section3.4} explains how we compute the quantitative performance metrics.  Note that Steps 2 and 3 correspond to, respectively, the first and second modeling components mentioned in Section~\ref{section1}. In order to provide an easy reference for the notations, Table~\ref{table2} provides a nomenclature list  for the important notations used in Sections~\ref{section3.1}--~\ref{section3.3}.
	\begin{table} [ht]
		\centering
		\caption{Nomenclature table.} \label{table2}
		\resizebox{\textwidth}{!}{
			\begin{tabular}{l  l}
				\hline
				Notation & Meaning\\
				\hline \hline
				$Q_1$ & Index set for the first dataset before covariate matching\\
				$Q_2$ & Index set for the second dataset before covariate matching\\
				$x_Q$ & Values of covariate $x$ in the index set Q\\
				$Q_A$ & Index set satisfying some condition $A$\\
				$\varpi$ & Thresholding coefficient regulating the tightness of the matching\\
				$D_1$ & First dataset after covariate matching\\
				$D_2$ & Second dataset after covariate matching\\
				$\mb{X}^{(1)}$ & $n_1 \times p$ matrix with input variable values for the first dataset after covariate matching\\
				$\mb{X}^{(2)}$ & $n_2 \times p$ matrix with input variable values for the second dataset after covariate matching\\
				$\boldsymbol{y}^{(1)}$ & Vector of length $n_1$ with output (power produced) for the first dataset after covariate matching\\
				$\boldsymbol{y}^{(2)}$ & Vector of length $n_2$ with output (power produced) for the second dataset after covariate matching\\
				$f_1(\cdot)$ & True underlying power curve for the first dataset \\
				$f_2(\cdot)$ & True underlying power curve for the second dataset \\
				$\hat{f}_1(\cdot)$ & Estimate of the power curve for the first dataset\\
				$\hat{f}_2(\cdot)$ & Estimate of the power curve for the second dataset\\
				$\mb{X}_{test}$ & $n_{test} \times p$ matrix of input points on which the functions $\hat{f}_1$ and $\hat{f}_2$ would be compared\\
				$\Delta_\text{unweighted}$ &  Difference between the estimated power curves on the test points $\mb{X}_{test}$\\
				$\Delta_\text{weighted}$ & Weighted difference between the estimated power curves on the test points $\mb{X}_{test}$\\\
				$\Delta_\text{scaled}$ & Difference between the estimated power curves scaled to the before-matching datasets \\
				\hline
		\end{tabular}}
	\end{table}
	
	\subsection{Subset selection and covariates ranking}\label{section3.1}
	This step is to screen the input covariates and select the covariate subset that can best explain the power output. The gold standard for subset selection is a randomized 5-fold or 10-fold cross-validation~\cite{Hastie2009}. The criterion used for selection is either a root mean squared error (RMSE) or a mean absolute error (MAE) \citep[Chapter 7]{Hastie2009}.  Using either criterion produces similar outcomes, as under a normality assumption for the errors, both RMSE and MAE would result in the same optimal value.  In this study, we use a 10-fold cross-validation and RMSE.
	
	In order to carry out the cross-validation, one would need to choose a regression method for modeling the power curve. Here we recommend using the k-nearest neighbors (kNN) method owing to its simplicity and computational efficiency \citep[Chapter 2]{Hastie2009}. However, practitioners are free to choose from a variety of data science models---an array of those are explained in Chapter 5 in~\cite{Ding2019}. When using different data science models, it is possible that a different best subset could have been selected.  When testing on the wind farm data at hand using the two best performing methods for power curve modeling---kNN and additive multiplicative kernel (AMK)---as per \cite[Chapter 5]{Ding2019}, we find that the subset selection outcome is rather insensitive to the choice of data science model used.
	
	The way to use the kNN model for establishing a power curve model is explained in Section 5.3.1 in~\cite{Ding2019}.  With the availability of packages implementing kNN in various programming languages like in \texttt{R}, \texttt{Python}, and \texttt{MATLAB}, a practitioner can simply arrange the data into training pairs, $\{\boldsymbol x_i, y_i, i=1,\ldots, n\}$, where $n$ is the number of the data points, and then utilize the appropriate function in one of the packages to fit a model.  The aforementioned cross-validation is an essential and standard procedure; see, for instance, Algorithm 2.1 in~\cite{Ding2019}.
	
	To select the best subset, we recommend using a greedy algorithm, either the forward stepwise selection or backward stepwise selection \citep[Chapter 3]{Hastie2009}. In our implementation we use the forward selection.  When we test the backward selection on our data, it does not produce a different outcome.
	
	The forward selection is to screen all the variables in $\boldsymbol x$, provisionally choose one variable at a time to build a model and compute the corresponding RMSE (or MAE) for that model.  Choose the variable that produces the smallest RMSE and add that to the model.  Conditioned on the variables that have been this far added, screen the remaining variables, which is to, again, provisionally add one variable at a time and choose the one that can produce the greatest reduction in RMSE.  The process stops when the resulting model's RMSE is no longer reduced on the addition of a variable. The subset of variables that produce the smallest RMSE is the best subset we are seeking and the order by which each variable is selected shows the natural order of covariate importance.
	
	The above procedure can be applied to every turbine on the farm to select the best subset for each of them.  Because the number of covariates in $\boldsymbol x$ needs to be the same when it gets to the later steps, one can create the union of the best subsets and use the union as the final $\boldsymbol x$.  Alternatively, one can select one or two representative turbines and use them to select the best subset for the whole farm; this latter approach is to shorten the time for selection.  Our experience indicates that the two approaches yield final outcomes that are not dissimilar.
	
	Using the latter approach, we select Turbines \#45 and \#54 as our representative turbines. The two turbines are used because their wind direction data are directly calibrated by the wind farm operator. Table~\ref{table3} presents the RMSEs as we go through the series of models having different number of covariates.  For both turbines, the three-variable subsets including $\{W, T, TI\}$ is the best subset.  We, therefore, use the three covariates in the subsequent analysis.  The importance order of the three covariates is the sequence in which they are selected, i.e., $W>T>TI$.
	
	\begin{table} [ht]
		\centering
		\caption{Subsect selection and importance ordering. RMSE values below are in percentage relative to the rated power. The best subset is shown in boldface font. } \label{table3}
		\begin{tabular}{lcc}
			\hline Covariates & \#45's RMSE & \#54's RMSE\\
			\hline \hline
			$W$ & 4.06\% & 3.08\% \\
			$W$, $T$ & 2.88\% & 2.42\% \\
			$\textit{\textbf W}$, $\textit{\textbf T}$, $\textit{\textbf{TI}}$ & \textbf{2.80}\% &\textbf{2.37\%}\\
			$W$, $T$, $TI$, $D$ &3.03\% & 2.68\%\\
			$W$, $T$, $TI$, $D$, $sdD$ & 3.27\% & 2.85\% \\
			\hline
		\end{tabular}
	\end{table}
	
	\subsection{Covariate matching}\label{section3.2}
	The method for covariate matching is explained in Section 7.2.1 in~\cite{Ding2019}. To state it formally,  let us adopt some notations from~\cite{Ding2019}. Denote the index set of the data records of two periods or two turbines by $Q_1$ and $Q_2$ respectively.  Here ``1'' and ``2'' bear generic meanings, not referring to Turbines \#1 and \#2 on the wind farm.  Let $x_{Q}$ denote the values of a covariate $x$ for data indices in $Q$. For example, $W_{Q_1}$ is the vector of wind speed values associated with Period/Turbine 1.  The attempt to make the wind speed in $Q_1$ to match data record $j$ in $Q_2$ can be expressed as
	
	\vspace{-12 pt}
	\begin{equation}\label{equ1}
		Q_{W_j} := \{i \in Q_1:\vert W_i-W_{j}\vert < \varpi \cdot \sigma(W_{Q_1})\},
	\end{equation}
	
	\noindent where $\sigma(x)$ is the standard deviation of $x$ in the specified dataset and $\varpi$ is the thresholding coefficient regulating the tightness of the matching. By carrying out this operation, we treat the data indexed by $Q_2$ as the baseline. If we repeat this operation for every data record in $Q_2$, then $Q_W$ is the index of the matched data records in $Q_1$.
	
	In the above matching process, two situations need to be addressed.  One situation is that $Q_{W_j}$ could turn out to be an empty set, meaning that no data point in $Q_1$ is found close enough to $W_j$ in $Q_2$.  When that happens, then let it be so.  The other situation is when $Q_{W_j}$ has multiple elements.  In this case, we recommend selecting the data point closest to the data record to be matched and putting the rest of the data points back in $Q_1$.
	
	The number of covariates involved in this step is the best covariate subset identified in the preceding step.  Since the best covariate subset contains multiple covariates (the chance that Step 1 selects only a single covariate is next to zero), we need to apply the matching procedure sequentially, i.e., through Algorithm 7.1 in~\cite{Ding2019}, which is labeled as \emph{hierarchical subgrouping}.  The default sequence of the hierarchical subgrouping is from the most important covariate to the lest important covariate.  This is the reason why we want to rank the importance of covariates in Section~\ref{section3.1}.  When there are multiple layers of matching, the thresholding coefficient, $\varpi$, can be different at each layer.  To make things simple, however, we recommend selecting a single constant threshold for the whole procedure. In our study, we set $\varpi=0.2$.
	
	There is one more technical point to be stressed.  The above procedure of covariate matching treats one of the periods/turbines as the baseline.  When the baseline is switched to the other period/turbine, there is a small discrepancy between the two data subsets selected.  In order for the subsequent comparison between two periods/turbines to be symmetric, meaning that regardless of which period/turbine is used as the baseline, the comparison outcome remains the same, a simple fix is to conduct the covariate matching twice for each pair-wise comparison. One uses Period/Turbine $i$ as the baseline and the other uses Period/Turbine $j$ as the baseline.  The two matched datasets are then combined, duplicates of any datapoints are eliminated, and the resulting dataset is used as the final matched dataset.
	
\begin{figure}
		\centering \includegraphics[width=.8\textwidth]{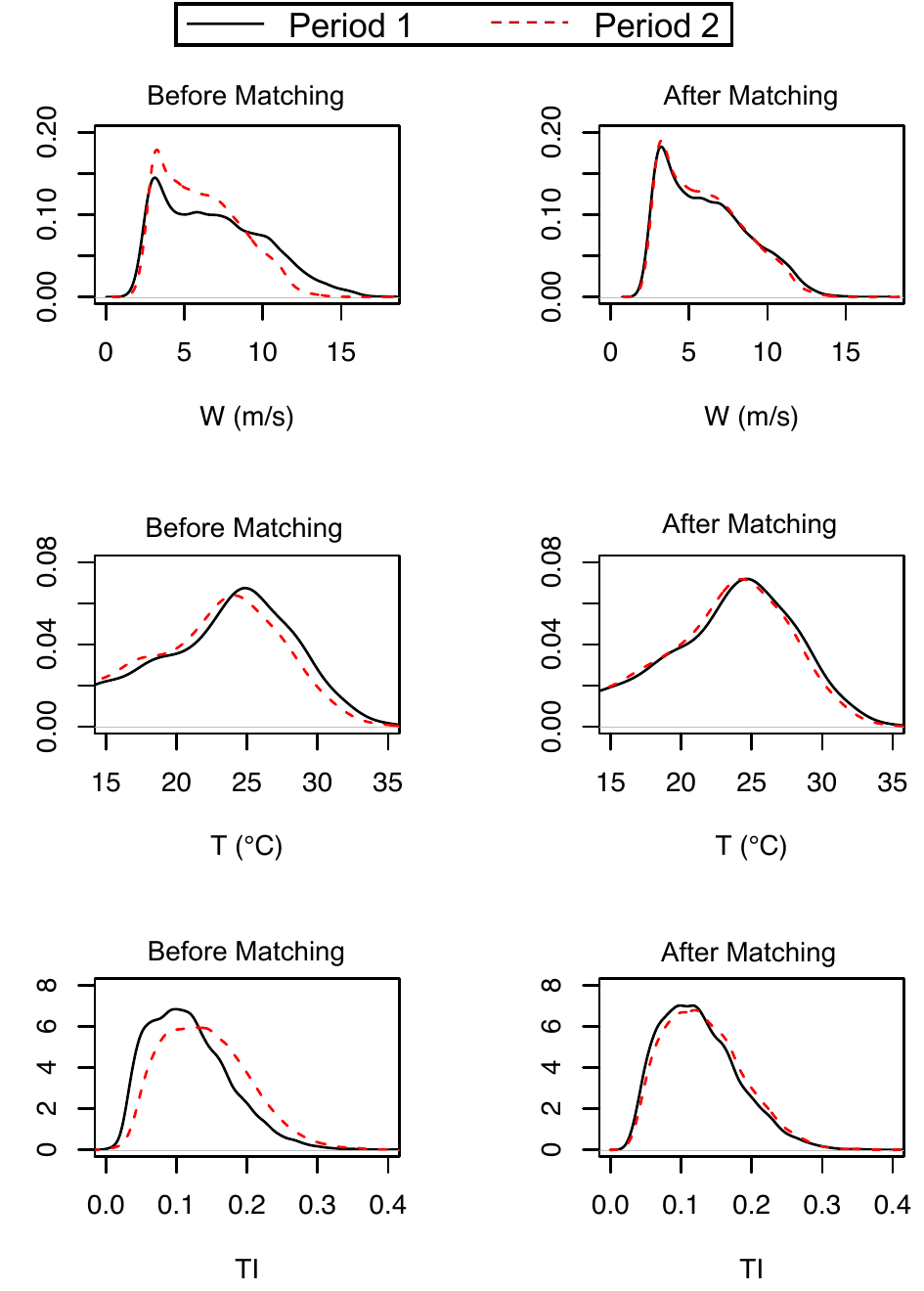}
		\caption{Probability density functions of the three selected covariates before and after covariate matching.}\label{figure3}
	\end{figure}

	Figure~\ref{figure3} illustrates what this covariate matching is meant to accomplish.  The left panel shows the empirically estimated probability density functions (pdf) before the matching, whereas the right panel shows the pdfs after the matching. Apparently, before the matching, the wind and environmental conditions of the two periods differ noticeably, whereas after the matching, they agree with each other considerably.
	
	\subsection{Gaussian process model for functional curve comparison}\label{section3.3}
	Prakash et al.~\cite{Prakash2020} propose a method for functional comparison, which tests the null hypothesis that the functions are equal at all input points. When the null hypothesis is rejected, Prakash et al.'s method identifies the region in the input space where the functions in comparison are different.  Their method is based on GP regression, a popular machine learning method~\cite{Rasmussen2006}, nonparametric in nature, and broadly used in various data science tasks.
	
	Recall that Section~\ref{section3.1} selects the covariate subset of dimension $p$, which, for our wind farm, are $W$, $T$, and, $TI$, whereas Section~\ref{section3.2} selects the statistically comparable data subsets for Turbine/Period 1 and 2, which we denote as $\{D_i, i = 1,2\}$ of $n_i$ data points each. Dataset $D_1$ can be denoted by an ordered pair $ \{\mathbf{X}^{(1)},\boldsymbol{y}^{(1)}\}$, where $ \mathbf{X}^{(1)} $ is an $ n_1 \times p $ matrix with each row corresponding to the covariates of one data point and $ \boldsymbol{y}^{(1)} $ is an $n_1\times 1$ vector with each element corresponding to the power output of the data pair. Similarly $D_2$ can be denoted as $ \{\mathbf{X}^{(2)},\boldsymbol{y}^{(2)} \} $.  Prakash et al.~\cite{Prakash2020} assume that these datasets come from underlying models given by:
	
	\vspace{-12 pt}
	\begin{equation}
		y_{ij} = f_i(\boldsymbol{x}_{ij}) + \epsilon_{ij}, \quad i = 1,2, \quad  j = 1,\dots,n_i,
	\end{equation}
	
	\noindent where $f_1(\cdot)$ and $f_2(\cdot)$ are  two smooth continuous functions with the same domain of $\mathcal{X} \subseteq \mathbb{R}^p$ and $\epsilon_{ij} \stackrel{iid}{\sim} \mathcal{N}(0,\sigma_{\epsilon}^2)$ with $ \sigma_{\epsilon}^2$ as the variance of the noise. Prakash et al.~\cite{Prakash2020} mean to test the following hypotheses, namely, under the null hypothesis $H_0$:
	
	\vspace{-12 pt}
	\begin{equation}
		f_1(\boldsymbol{x}) = f_2(\boldsymbol{x}) \quad \text{for} \ \boldsymbol{x} \in \mathcal{X},
	\end{equation}
	
	\noindent and under the alternative hypothesis, $H_1$:
	
	\vspace{-12 pt}
	\begin{equation}
		\text{there exists an} \,\, \boldsymbol{x} \in \mathcal{X} \quad s.t. \quad f_1(\boldsymbol{x}) \neq f_2(\boldsymbol{x}).
	\end{equation}
	
	For the wind turbine application, $f_1(\cdot)$ and $f_2(\cdot)$ are the multi-dimensional power curve functions associated with Period/Turbine 1 and 2, respectively.  The true $f_1(\cdot)$ and $f_2(\cdot)$ are, of course, unknown and they are to be estimated from the data. To do that, Prakash et al.~\cite{Prakash2020} assume that $f_1(\cdot)$ and $f_2(\cdot)$ are samples from a GP with zero mean and a covariance function given by $ k(\boldsymbol{x},\boldsymbol{x}')$.
	
	Then, define a cross-covariance matrix $ \mathbf{K(X,X')} $ between a pair of input variable matrices, $ \mathbf{X} $ and $ \mathbf{X}' $, and a covariance vector, $ \boldsymbol{r}(\boldsymbol{x}) $, between the input data $ \mathbf{X} $ and any point $ \boldsymbol{x} $ as follows:
	
	\vspace{-12 pt}
	\begin{eqnarray}\label{CovMatNotations}
		\mathbf{K(X,X')} =
		\begin{bmatrix}
			k(\boldsymbol{x}_1,\boldsymbol{x}'_1) & k(\boldsymbol{x}_1,\boldsymbol{x}'_2) & \dots & k(\boldsymbol{x}_1,\boldsymbol{x}'_{n'})\\
			k(\boldsymbol{x}_2,\boldsymbol{x}'_1) & k(\boldsymbol{x}_2,\boldsymbol{x}'_2) &\dots  & k(\boldsymbol{x}_2,\boldsymbol{x}'_{n'})\\
			\vdots & \vdots & \ddots & \vdots \\
			k(\boldsymbol{x}_n,\boldsymbol{x}'_1) & k(\boldsymbol{x}_{n},\boldsymbol{x}'_2) & \dots & k(\boldsymbol{x}_{n},\boldsymbol{x}'_{n'})\\
		\end{bmatrix},
		\quad
		\boldsymbol{r}(\boldsymbol{x}) =
		\begin{bmatrix}
			k(\boldsymbol{x}_1,\boldsymbol{x})\\
			k(\boldsymbol{x}_2,\boldsymbol{x})\\
			\vdots\\
			k(\boldsymbol{x}_{n},\boldsymbol{x})\\
		\end{bmatrix},
	\end{eqnarray}
	
	\noindent where $ \boldsymbol{x}_1 \dots \boldsymbol{x}_{n}$ are the vectors in the rows of $ \mathbf{X} $, and $ \boldsymbol{x}'_1 \dots \boldsymbol{x}'_{n'} $ are the vectors in the rows of $ \mathbf{X}'$. The estimated power curve function for $f_1(\cdot)$ given  $D_1$, and that for $ f_2(\cdot) $ conditioned on  $D_2$, are:
	
	\vspace{-12 pt}
	\begin{equation}\label{posteroirFn}
		\begin{split}
			\hat{f}_1(\boldsymbol{x}) &= \boldsymbol{r}_1(\boldsymbol{x})^\top[\mathbf{K}(\mb{X}^{(1)},\mb{X}^{(1)})+\sigma_{\epsilon}^2\mathbf{I}_{n_1}]^{-1}\boldsymbol{y}^{(1)}, \\
			\hat{f}_2(\boldsymbol{x}) &= \boldsymbol{r}_2(\boldsymbol{x})^\top[\mathbf{K}(\mb{X}^{(2)},\mb{X}^{(2)})+\sigma_{\epsilon}^2\mathbf{I}_{n_2}]^{-1}\boldsymbol{y}^{(2)},
		\end{split}
	\end{equation}
	
	\noindent where $ \boldsymbol{r}_i(\boldsymbol{x}) $, for $i=1$ or $2$, is the covariance vector between $ \mathbf{X}^{(i)} $ and any point $ \boldsymbol{x} $, and $ \mathbf{I}_{n_i} $ is an $n_i\times n_i$ identity matrix.  The estimated power curve functions given in Equation~\eqref{posteroirFn} are the best linear unbiased predictors for $ f_1(\cdot) $ and $ f_2(\cdot)$, respectively, under the GP model~\cite{Rasmussen2006}.
	
	Once the two power curve functions are estimated, we can test the function comparison hypothesis on a dense regular grid $\mathcal{T} \subset \mathcal{X}$. Assume that the number of test points in  $\mathcal{T}$ is $n_{test}$ and define a matrix $ \mathbf{X}_{test} $ such that each row of the matrix represents one test point in $ \mathcal{T} $.  Then, we can use the above power curve functions to predict power output for all the points on the grid of  $\mathcal{T}$, as follows:
	
	\vspace{-12 pt}
	\begin{equation}\label{Eqn:posterior1}
		\begin{split}
			\boldsymbol{\hat{f}}_1 & =\mathbf{K}(\mb{X}_{test},\mb{X}^{(1)})[\mathbf{K}(\mb{X}^{(1)},\mb{X}^{(1)})+\sigma_{\epsilon}^2\mathbf{I}_{n_1}]^{-1}\boldsymbol{y}^{(1)},\\
			\boldsymbol{\hat{f}}_2 & =\mathbf{K}(\mb{X}_{test},\mb{X}^{(2)})[\mathbf{K}(\mb{X}^{(2)},\mb{X}^{(2)})+\sigma_{\epsilon}^2\mathbf{I}_{n_2}]^{-1}\boldsymbol{y}^{(2)},
		\end{split}
	\end{equation}
	
	\noindent where $ \boldsymbol{\hat{f}}_1  = (\hat{f}_1(\boldsymbol{x}_1) \dots \hat{f}_1(\boldsymbol{x}_{n_{test}}) )^\top  $, and $ \boldsymbol{\hat{f}}_2  = (\hat{f}_2(\boldsymbol{x}_1) \dots \hat{f}_2(\boldsymbol{x}_{n_{test}}) )^\top  $.  The hypothesis test is then to test whether $\boldsymbol{\hat{f}}_2 -  \boldsymbol{\hat{f}}_1 = \mathbf 0$ or not.
	
	In our application, we decide to use the two most important factors identified in Section~\ref{section3.1} to fit the GP-based power curve model and conduct the testing.  The two factors are $W$ and $T$. The reason we use the two-variable input is to make the computation in the functional comparison more tractable.  From Table~\ref{table3}, it is clear that when using a three-variable power curve model, it only improves the model accuracy marginally (about 2--3\%) from the two-variable model, while including $T$ improves the model accuracy substantially (22--30\%) as compared with the one-variable model having only $W$.
	
\begin{figure}
		\centering \includegraphics[width=\textwidth]{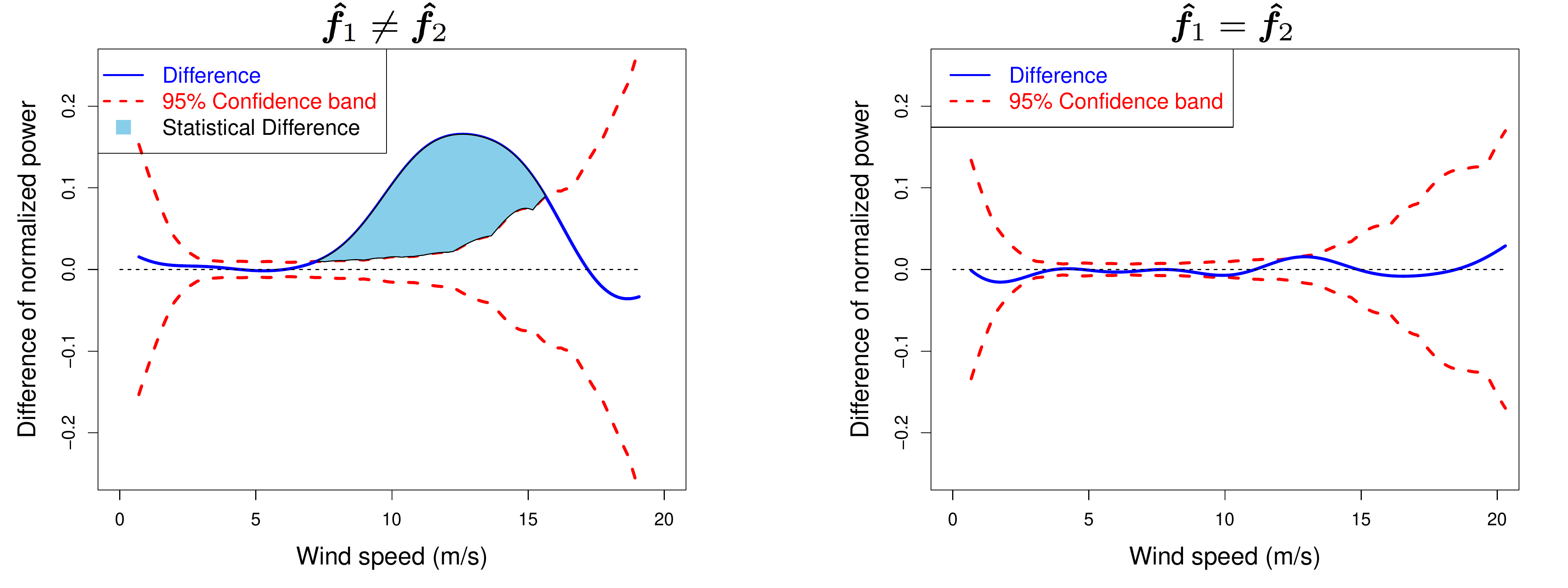}
		\caption{Power difference curves and the corresponding $95\%$ confidence interval.}\label{figure4}
	\end{figure}

	Figure~\ref{figure4} illustrates the outcomes of this GP-based functional curve comparison. This example uses the data from Turbine \#41.  The two panels plot the power difference curve (i.e., $\boldsymbol{\hat{f}}_2 -  \boldsymbol{\hat{f}}_1$) versus the wind speed, as well as the associated $95\%$ confidence band.  When a portion of the power difference curve is outside the $95\%$ band, we can say that the null hypothesis, $\boldsymbol{\hat{f}}_2 -  \boldsymbol{\hat{f}}_1 = \mathbf 0$, is rejected at the significance level of $95\%$.
	
	The left panel is the power difference curve between \#41's period 2 versus period 3 and the null hypothesis is rejected. When a rejection happens, this power difference curve and its $95\%$ band clearly identifies the region of rejection.  In this case, it is from 7 m/s to about 17 m/s, i.e., the Region II and part of Region III on a power curve, which are important for power production. For now, one can ignore the statistical difference, which will be explained in the next subsection.
	
	The right panel is from \#41's period 3 but the data of that period is split into two subsets via random sampling. And the comparison of these two subsets within a single period produces a power difference curve close to zero at most of the input domain, except at the high wind speed end.  At the high wind speed end, the uncertainty is high, as evidenced by a much broader 95\% confidence band.  This is due to the scarcity of data in that high wind region.  The entire power difference curve is inside the 95\% confidence band, so that the null hypothesis cannot be rejected, suggesting that there is no strong enough evidence, up to $95\%$ level of significance, to state that the power curves associated with these two subsets of data are different.
	
	\subsection{Performance metrics}\label{section3.4}
	To quantify the difference between two turbines or two periods, a power difference metric in the unit of kilo-Watts (kW) is defined for the two turbines/periods involved, as follows:
	
	\vspace{-12 pt}
	\begin{equation}\label{equ8}
		\Delta = \frac{1}{n_\text{test}}\sum_{i=1}^{n_\text{test}} (\hat{f}_2(\boldsymbol{x}_i)-\hat{f}_1(\boldsymbol{x}_i)).
	\end{equation}
	
	\noindent Then, the power difference, in percentage, is defined as
	
	\vspace{-12 pt}
	\[ \Delta\% = \frac{\Delta}{\frac{1}{n_\text{test}}\sum_{i=1}^{n_\text{test}}\hat{f}_1(\boldsymbol{x}_i)}\times 100\%. \]
	
	\noindent Here we use Turbine 1 or Period 1 as the base while computing the percentage.  This undertaking is based on the advisement of our collaborator to be consistent with the common practice in industry.  Practitioners who desire to use Turbine 2 or Period 2 as the base can do so by simply replacing the denominator with $\frac{1}{n_\text{test}}\sum_{i=1}^{n_\text{test}}\hat{f}_2(\boldsymbol{x}_i)$.
	
	The above $\Delta$ and its percentage are unweighted as it is the simple average of the power responses on all testing inputs. Let us label it as $\Delta_\text{unweighted}$ and the percentage as $\Delta\%_\text{unweighted}$.  In other words, the unweighted difference treats different $\boldsymbol x$'s as if they appear at the same frequency, but in actual turbine operations, different $\boldsymbol x$'s do not appear at the same frequency.  Recall an $\boldsymbol x$ is a combination of the wind and other environmental conditions.  Suppose $\boldsymbol x=(W, \,\, T)$.  Then the condition of $W=7$ m/s and $T=25^\circ$C appears over a year's period at a different frequency than some other condition like $W=18$ m/s and $T=-5^\circ$C.
	
	To account for the uneven distribution of $\boldsymbol x$'s, one should weigh $\Delta$  and $\Delta\%$ by the relative frequency associated with a specific $\boldsymbol x_i$, which is denoted by $P_i$ and can be estimated by counting the number of actual weather data in the original datasets falling into a small neighborhood around $\boldsymbol x_i$.  With $P_i$, the weighted power difference is
	
	\vspace{-12 pt}
	\begin{equation}\label{equ9}
		\Delta_\text{weighted} = \sum_{i=1}^{n_\text{test}} P_i\times(\hat{f}_2(\boldsymbol{x}_i)-\hat{f}_1(\boldsymbol{x}_i)).
	\end{equation}
	
	\noindent Then, the weighted power difference in percentage is
	
	\vspace{-12 pt}
	\[ \Delta\%_\text{weighted} = \frac{\Delta_\text{weighted}}{\sum_{i=1}^{n_\text{test}}P_i\times \hat{f}_1(\boldsymbol{x}_i)}\times 100\%. \]
	
	One of the benefits of using the GP-based functional curve comparison method, as described in Section~\ref{section3.3}, is that it provides uncertainty quantification in the form of a $100(1-\alpha)\%$ confidence band for the difference curve; please refer to Figure~\ref{figure4} for an illustration.  This prompts us to propose a new metric, labeled as \emph{statistical significant difference}, which is to accumulate only the portion that is beyond the $100(1-\alpha)\%$ confidence band, when counting the difference. The shaded portion in the left panel of Figure~\ref{figure4} corresponds to the statistical significant difference. The use of statistical significant difference is to make the quantification robust by weeding out the portions that may have been caused by random fluctuation in data.  The statistical significant difference can be computed also in the form of either unweighted or weighted, as explained above.
	
	If the difference curve is entirely within the confidence band, then the statistical significant difference between the two cases is zero. Please refer to the right panel of Figure~\ref{figure4} for such an example.
	
	So far, these quantification measures are computed using the subset of data after the covariate matching. In order to scale the difference back to the original dataset, one approach commonly used in industry is to reweigh the difference by using the power distribution of the original data.  The specific steps are as follows:
	\begin{enumerate}
		\item Partition the power spectrum into $K$ bins.  The default we use is $K=15$, which, for a 1.5MW class turbine, translates to a bin width of 100 kW.
		\item Use $\hat{f}_1(\cdot)$ as the reference. Find the $\boldsymbol x$'s, for which the corresponding $\hat{f}_1(\boldsymbol x)$ falls into power bin $k$. Record these $\boldsymbol x$'s in the set of $Q_k$.
		\item Compute
		\begin{equation*}
			\begin{split}
				\mu_k  &=\text{avg}_{\boldsymbol x \in Q_k}\left(\hat{f}_1(\boldsymbol x)\right), \\
				\delta_{k} &= \text{avg}_{\boldsymbol x \in Q_k}\left(\hat{f}_2(\boldsymbol x)-\hat{f}_1(\boldsymbol x)\right).
			\end{split}
		\end{equation*}
		\item Use the original data of both periods (or both turbines) to compute the relative frequency (histogram) of each power bin and denote the relative frequency by $\pi_k$, for $k=1, \ldots, K$.
		\item The difference metric between two turbines/periods, scaled back to the original dataset, is then calculated by
		
		\vspace{-12 pt}
		\begin{equation}
			\Delta_\text{scaled} = \sum_{k=1}^{K} \pi_k \cdot \delta_k.
		\end{equation}
		
		\noindent Then, the scaled power difference in percentage is
		
		\vspace{-12 pt}
		\[ \Delta\%_\text{scaled} = \frac{\Delta_\text{scaled}}{\sum_{k=1}^{K}\pi_k \mu_k} \times 100\%. \]
		
	\end{enumerate}
	
	Please note that although the power difference in Steps 2 and 3 are computed in an unweighted fashion, like in Equation~\ref{equ8}, the final scaled metric is weighted by the power spectrum. For this reason, the scaled power difference is a weighted metric more like in Equation~\ref{equ9}.

	\section{Analysis and Discussion}\label{section4}
	
	\subsection{Application to known upgrades studied before}\label{section4.1}
	Before we apply the proposed evaluation method to the case of 66 wind turbines, we would like to verify the credibility of the method.  We choose to apply the method to the two datasets of turbine upgrades studied before: one is a simulated case from pitch angle adjustment, and the other is a real, physical modification of VG installation; see~\cite{Lee2015a}, \cite{Shin2018}, and also \cite[Chapter 7]{Ding2019}.  Each dataset has two turbine subsets---one is for the test turbine on which a modification was made, whereas the other is a control turbine on which no modification was made. The datasets are available and retrieved from \url{https://aml.engr.tamu.edu/book-dswe/}, the \texttt{Turbine Upgrade Dataset}. A detailed explanation of the datasets can also be found on the same website.
	
	The pitch angle adjustment is a simulated case, in the sense that the increase to the after-upgrade power was added artificially.  Specifically, the wind power in the after-upgrade section is multiplied by $(1+r)$ for those whose corresponding wind speed is 9 m/s or above, where $r$ takes the value of $0.02$, $0.03$, $\ldots$, $0.09$.  This process is to simulate a power increase to a subset of wind power by $2\%$ to $9\%$, respectively.  The effective increase to the after-upgrade portion is not $r$, as the wind power whose corresponding wind speeds are smaller than 9 m/s are unaffected.  The effective increase in power is denoted by $r^\prime$ in \cite[Chapter 7]{Ding2019} and defined in Equation~(7.7) therein.
	
	The purpose of analyzing this simulated data, as explained in the previous studies, is to assess the capability of an evaluation method in the context of a broad range of small to moderate increases in turbine performance. For a simulated case, the underlying true increase is known, so that such assessment can be conducted. By contrast, for a physical VG installation, the underlying true change is not known. Even as of now, no physically controlled experiment can be conducted on a commercial-size turbine to unearth the true benefit of a VG installation.  All the evaluation methods can only be compared against each other.
	
	We follow the procedure outlined in Section~\ref{section3} to analyze both upgrade cases. Step 1 identifies five factors with their importance order as $W>D>\rho>S>TI$, where $\rho$ is air density and $S$ is wind shear.  Step 2 uses all five of them and Step 3 uses the most important two, which are $W$ and $D$. To calculate the performance difference due to an upgrade, we use $\Delta_\text{weighted}$, rather than $\Delta_\text{scaled}$, because what the other methods calculated is effectively the weighted difference. Since each case has two datasets, the test set and the control set, the two sets are analyzed separately on their own.  For each set, the analysis is a temporal comparison, i.e., the change for that turbine before and after the upgrade time point.  Note that the upgrade time point is specified for each upgrade case in the \texttt{Turbine Upgrade Dataset}.
	
	Although the control turbine does not undergo a purposeful modification by the owner's engineering team, there is no guarantee that there is no change to that turbine either. So the final performance change to the test turbine is deemed as its own performance change, less that of the control turbine.  This is a common practice used by nearly all similar studies, e.g., in~\cite{Lee2015a}. In~\cite{Shin2018}, the spirit of this control-test difference is used but the mechanism for using it is slightly different. The method in Shin et al.~\cite{Shin2018} is similar to the second step of our proposed procedure in Section~\ref{section3}, i.e., the covariate matching.  In order to account for the natural change in a control turbine, Shin et al.~\cite{Shin2018} uses the power output of the control turbine as one of the covariates in its matching procedure.  Because of this, Shin et al.~\cite{Shin2018} has to be applied to a pair of control-test datasets and cannot be separately used on an individual turbine.  This complication is not very restrictive in the context of VG upgrade quantification because most of those VG cases in the literature use a pair of turbines anyway, but it does restrict the flexibility for general turbine performance analysis.
	
	Table~\ref{table4} presents the comparison of the proposed method with the other existing methods.  The proposed method is uniformly better than the method in~\cite{Lee2015a}.  The method in~\cite{Shin2018} does have its merit and for some of the cases, it estimates the upgrade more accurately.  But the advantage of the proposed method is its robustness.  Over the wide range of performance change, the estimate resulting from the proposed method is within 10\% of the target value.  That cannot be said of the method in~\cite{Shin2018}, as for some cases, the estimate from \cite{Shin2018} can be as bad as 39\% off the target value.  Another advantage of the proposed method, as we have explained earlier, is that it has the flexibility to be applied to individual turbines on their own, whereas the method in~\cite{Shin2018} always needs a pair of turbines to work.
	
	\begin{table}
		\caption{Comparison of three methods over different degrees of turbine performance change in the case of pitch angle adjustment. In the table, our estimate is $\Delta_\text{weighted}(\text{test})- \Delta_\text{weighted}(\text{control})$, where $\Delta_\text{weighted}(\text{control})=-0.07\%$ for all $r$ values. The \texttt{UPG} and \texttt{DIFF} are taken directly from~\cite{Ding2019}. }\label{table4}
		\resizebox{\textwidth}{!}{
			\begin{tabular}{lrrrrrrrrr}
				\hline\hline
				$r$     & 2\%  & 3\%  & 4\%  & 5\%  & 6\%  & 7\%  & 8\%  & 9\% \\
				$r^\prime$    & 1.25\% & 1.87\% & 2.49\% & 3.11\% & 3.74\% & 4.36\% & 4.98\% & 5.60\% \\
				\hline
				\tabularnewline
				\text{Our estimate} & 1.12\% & 1.77\% & 2.73\% & 3.43\% & 3.69\% & 4.69\% & 5.16\% & 5.59\% \\
				\text{Our estimate}/$r^\prime$ & \multicolumn{1}{l}{0.90} & \multicolumn{1}{l}{0.95} & \multicolumn{1}{l}{1.10} & \multicolumn{1}{l}{1.10} & \multicolumn{1}{l}{0.99} & \multicolumn{1}{l}{1.08} & \multicolumn{1}{l}{1.04} & \multicolumn{1}{l}{1.00} \\
				\hline
				\tabularnewline
				\text{UPG} in~\cite{Shin2018}& 1.74\% &2.21\% & 2.68\% & 3.16\% & 3.63\% & 4.11\% & 4.58\% & 5.05\% \\
				\text{UPG}/$r^\prime$ & \multicolumn{1}{l}{1.39}	& \multicolumn{1}{l}{1.18}	& \multicolumn{1}{l}{1.08} &	\multicolumn{1}{l}{1.02} &	\multicolumn{1}{l}{0.97}	& \multicolumn{1}{l}{0.94}	& \multicolumn{1}{l}{0.92}	& \multicolumn{1}{l}{0.90}  \\
				\hline\tabularnewline
				\text{DIFF} in~\cite{Lee2015a} & 1.97\% &2.56\% & 3.15\% & 3.73\% & 4.30\% & 4.86\% & 5.42\% & 5.97\% \\
				\text{DIFF}/$r^\prime$ & \multicolumn{1}{l}{1.58}	& \multicolumn{1}{l}{1.37}	& \multicolumn{1}{l}{1.27} &	\multicolumn{1}{l}{1.20} &	\multicolumn{1}{l}{1.15}	& \multicolumn{1}{l}{1.11}	& \multicolumn{1}{l}{1.09}	& \multicolumn{1}{l}{1.07}  \\
				\hline\hline
		\end{tabular}}
	\end{table}
	
	When the proposed method is applied to the VG upgrade pair, the resulting $\Delta_{\text{weighted}}(\text{test})= 2.29\%$ and $\Delta_{\text{weighted}}(\text{control})= 0.97\%$, which means our estimate of the turbine change due to the VG installation is the difference between these two values, i.e., $1.32\%$.
	
	The difficulty of the VG upgrade case is that we do not know the underlying truth. So we should put our estimate in the context of other methods.  According to~\cite[Chapter 7]{Ding2019}, the method in~\cite{Shin2018} estimates the change to be $1.13\%$ and the method in~\cite{Lee2015a} estimates the change to be $1.48\%$. Our new estimate is in between the previous two estimates, so we take that as a good sign.  The analysis in~\cite{Hwangbo2017b} also stated that the method in~\cite{Lee2015a} tends to overestimate, yet our current estimate is smaller than the estimate done by~\cite{Lee2015a}; we take this as another good sign.
	
	\subsection{Evaluate the technical upgrades}\label{section4.2}
	Next we apply the proposed method to the 66 turbine data explained in Section~\ref{section2}.  The first analysis is a temporal evaluation, which is to quantify each turbine's performance change for each of the three technical upgrades mentioned in Section~\ref{section2}. This analysis is carried out on all 66 turbines, so that for each pair of periods, there are 66 performance differences.  The 66 differences are plotted in a boxplot to give a sense of variation on the wind farm.
	
\begin{figure}
		\centering \includegraphics[width=\textwidth]{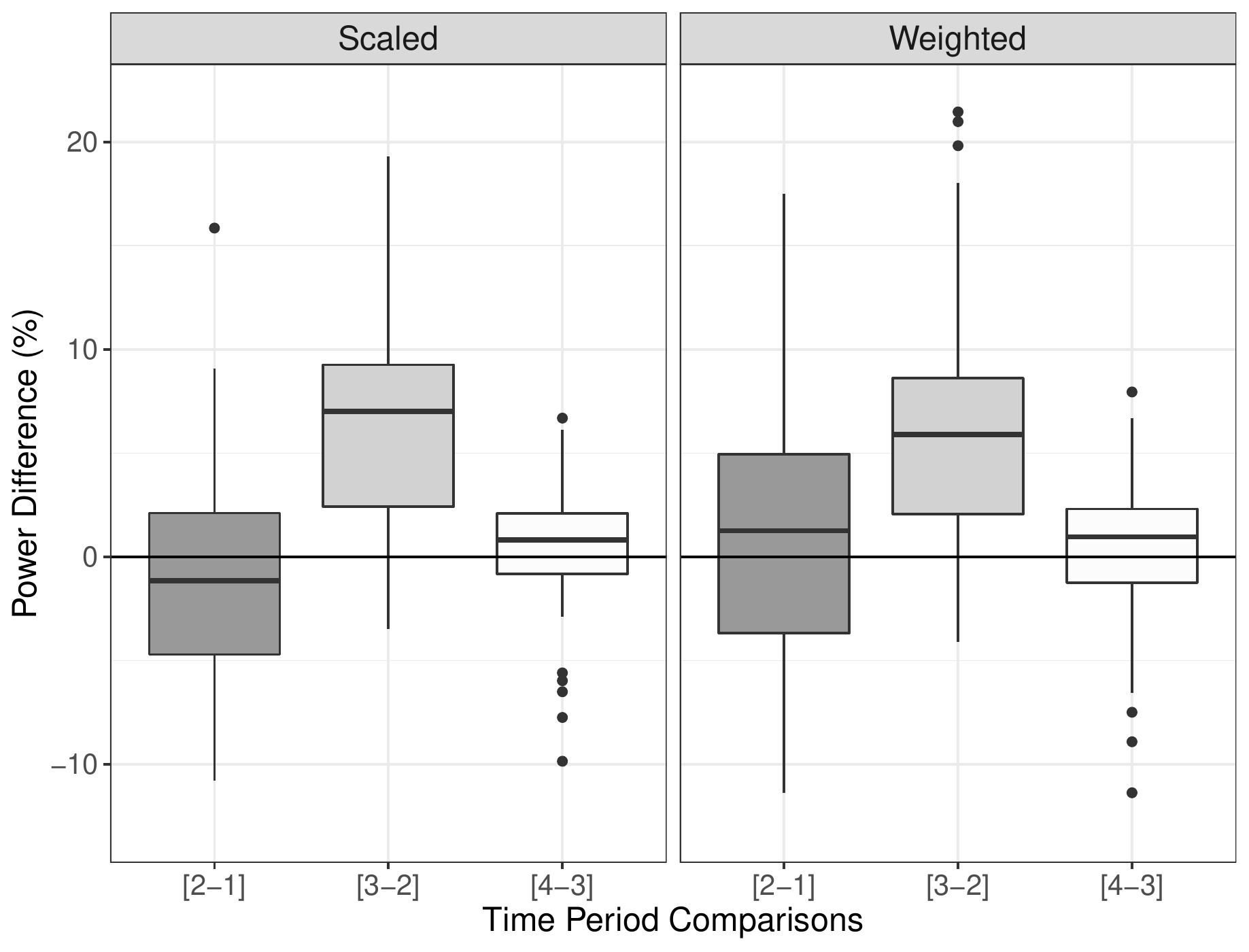}
		\caption{Boxplots of absolute power differences of the technical upgrades. Left panel: the scaled power difference $\Delta_\text{scaled}$; right panel: the weighted power difference $\Delta_\text{weighted}$.}\label{figure5}
	\end{figure}

	Figure~\ref{figure5} presents the boxplots of the absolute power difference (as opposed to statistical significant difference) for both $\Delta_\text{scaled}$ (left) and $\Delta_\text{weighted}$ (right).  It seems that the second technical action, V1, plays a rather substantially positive role in improving the turbines' performance, whereas the effect of the other two technical actions are not that significant or could even be detrimental (in the case of P1).  The average improvement made by V1 for the turbines is about a 6\% increase in their weather-weighted power production capability or an 8\% if scaled back to the original data.  It is also noted that the scaling to the original data does play a role in adjusting the power difference quantification.  The scaling seems to broaden the spread of the average $\Delta$'s.  Before the scaling, the average $\Delta$ ranges from $1\%$ to $6\%$, whereas after the scaling, the range becomes $-1\%$ to $8\%$.
	
\begin{figure}
		\centering \includegraphics[width=\textwidth]{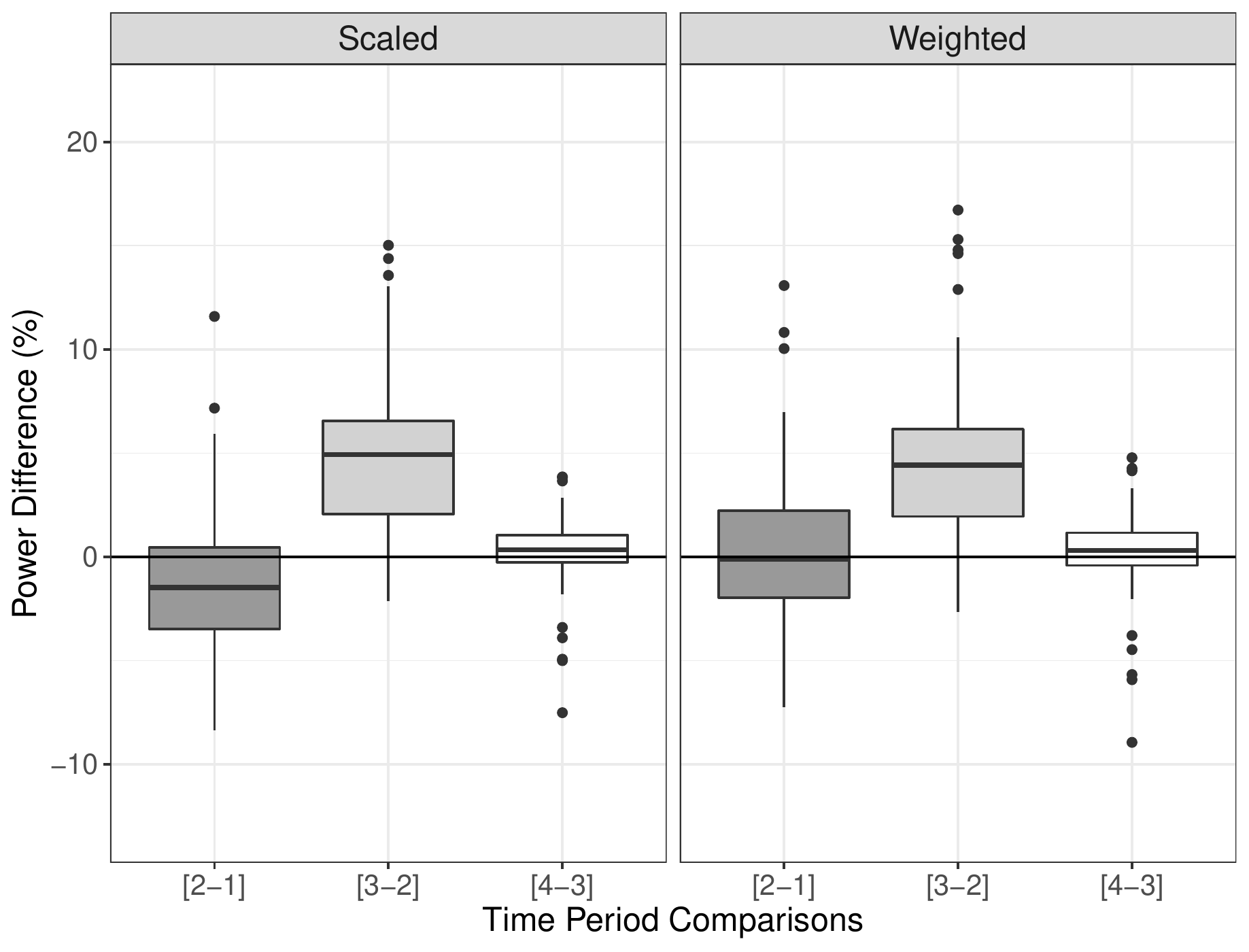}
		\caption{Boxplots of statistical significant power differences of the technical upgrades. Left panel: the scaled power difference $\Delta_\text{scaled}$; right panel: the weighted power difference $\Delta_\text{weighted}$.}\label{figure6}
	\end{figure}

	Figure~\ref{figure6} presents the boxplots in terms of statistical significant difference for both scaled (left) and weighted (right) scenarios. The message from the statistical significant different plot stays the same as in Figure~\ref{figure5} but the benefit of using the statistical significant plot is that the message becomes clear, as the variation becomes much smaller.  The average improvement made by V1 for the turbines is now about a 5\% increase in the scaled power production capability, 3\% smaller than the counterpart value in Figure~\ref{figure5}.  Loosely speaking, this 3\% may be attributed to noise fluctuation.
	
	\subsection{Space-time analysis}\label{section4.3}
	The second analysis is a space-time analysis, which compares a turbine with other turbines as well as a turbine with itself over the time. For this purpose, we need to select a baseline turbine, against which all other turbines are compared.  We decide to select Turbine \#12 as the baseline turbine, because it is the closest to the met tower.  We want to note that being close to the met tower does not play any significance in our analysis, so that this choice is a bit arbitrary.  We could possibly choose a different turbine as the baseline turbine as well.
	
\begin{figure}
		\subfigure[2015]{\includegraphics[width=0.5\textwidth]{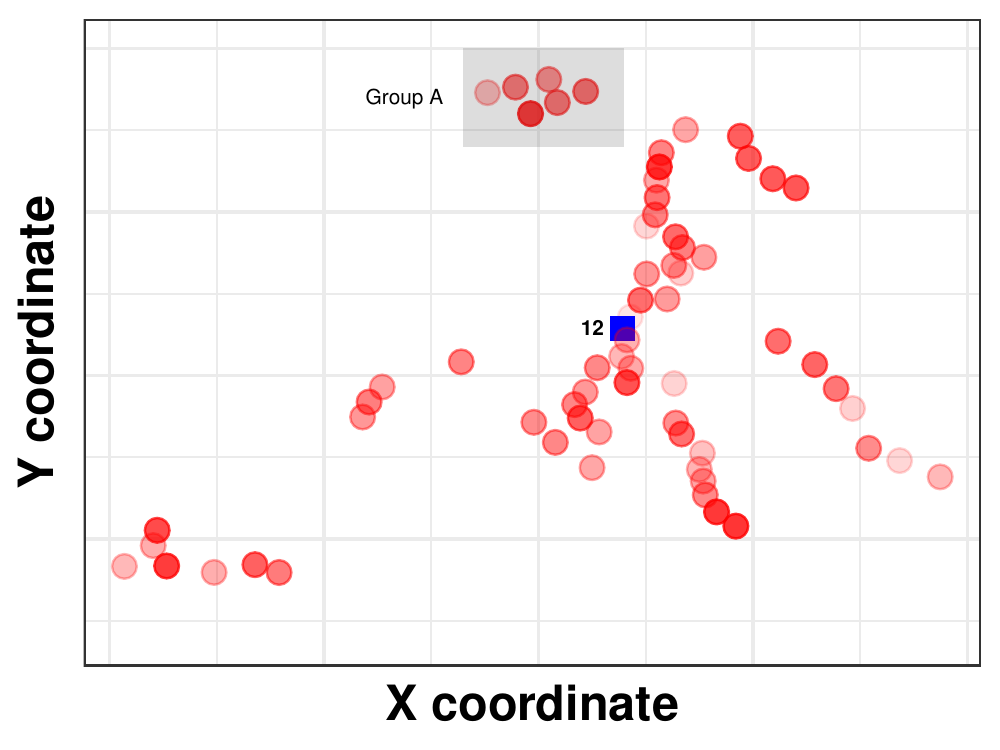}}
		\subfigure[2016]{\includegraphics[width=0.5\textwidth]{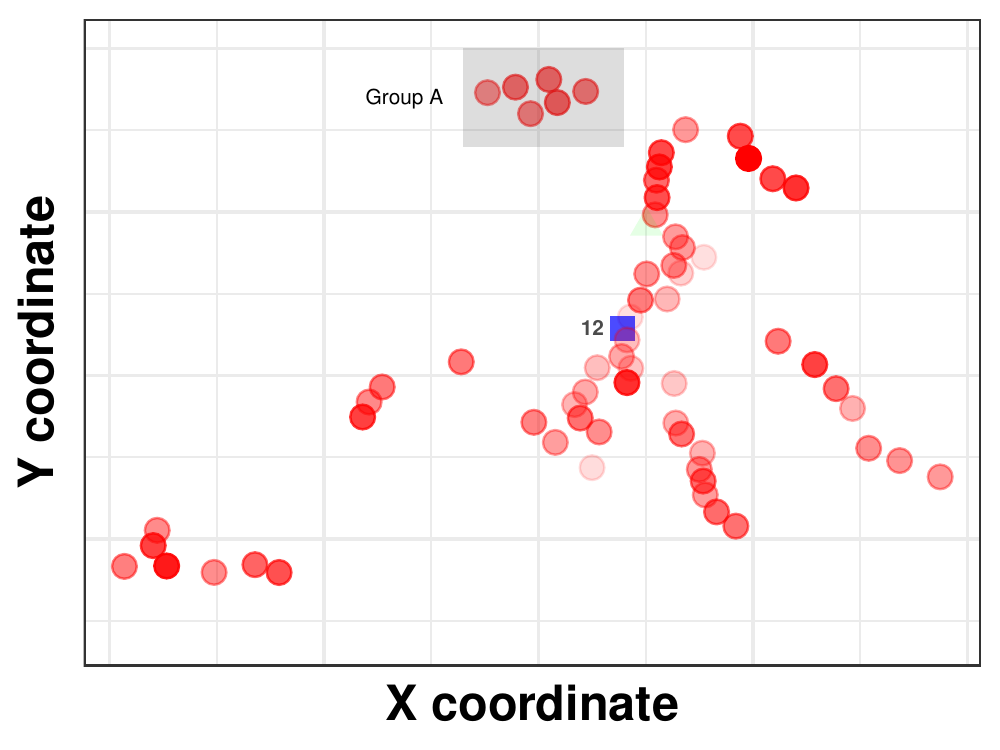}}
		\subfigure[2017]{\includegraphics[width=0.5\textwidth]{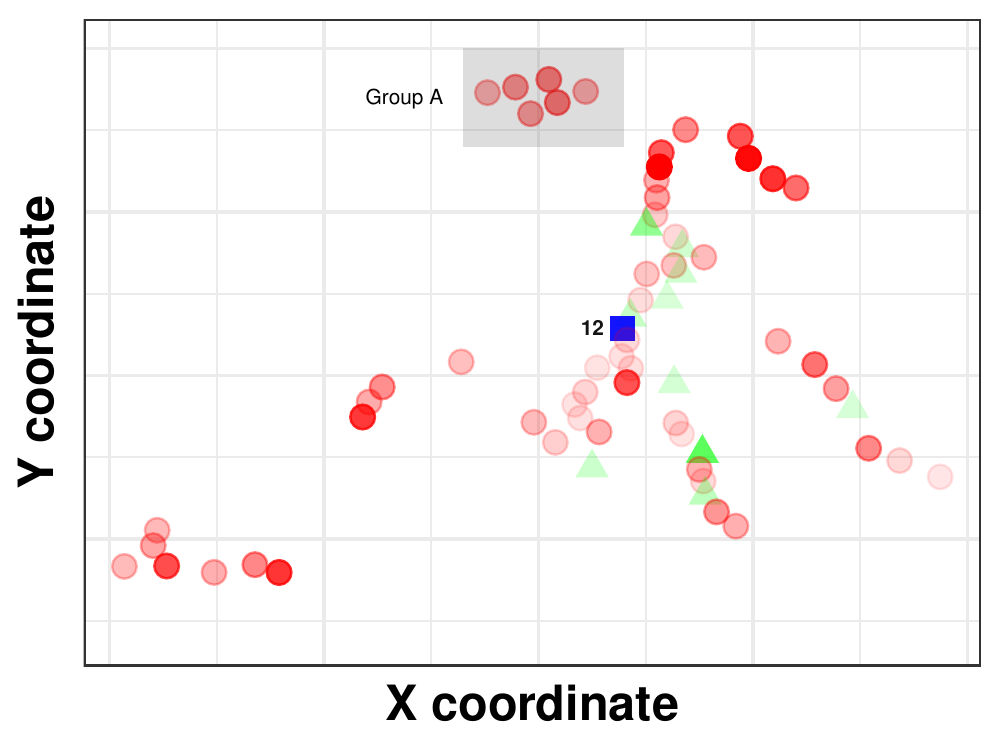}}
		\subfigure[2018]{\includegraphics[width=0.5\textwidth]{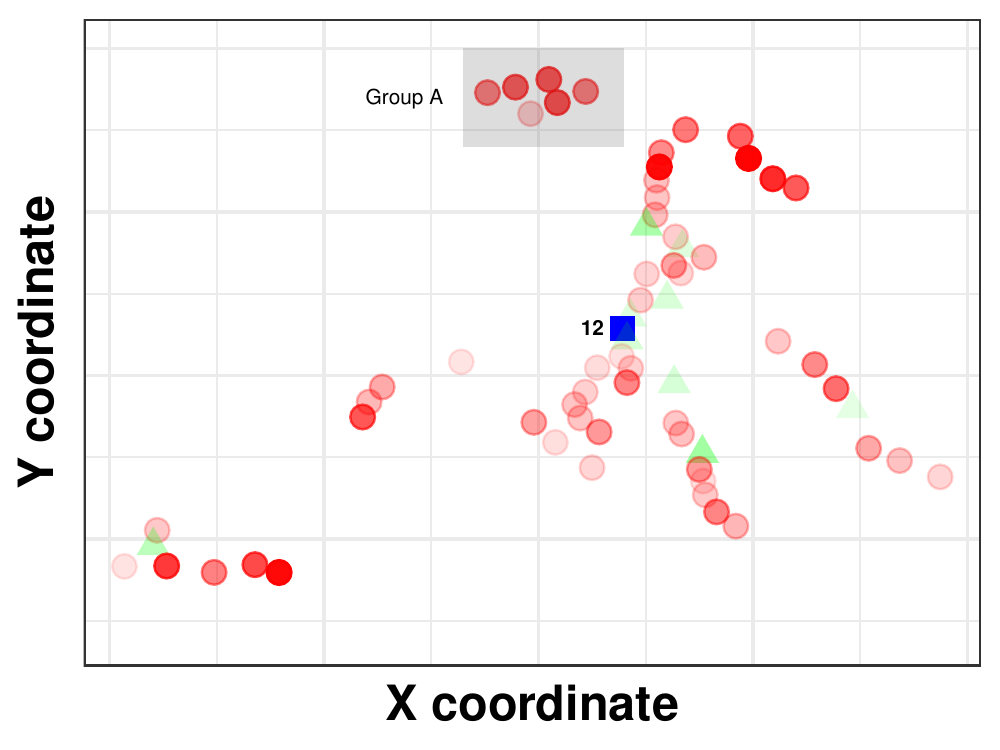}}
		\caption{Farm-wise performance comparison of wind turbines with Turbine \#12 as the baseline.}\label{figure7}
	\end{figure}

	Figure~\ref{figure7} is the comparison using the data of Turbine \#12 of each respective year as the baseline.  This comparison is technically only a space-wise comparison, because the baseline is reset for each calendar year.  A turbine shown as a red circle performs worse relative to \#12, while a turbine shown as a green triangle performs better relative to \#12. The color intensity is set to be proportional to the absolute value of the difference. Turbine \#12, itself, is marked as the blue square.
	
	Compared with \#12 in each year, it appears that initially almost all turbines perform worse relative to \#12.  In the last two years, 2017 and 2018, a subset of turbines perform better than \#12. This is most likely due to the three technical actions.  Although two of the three technical actions do not produce a significant change to the turbines \emph{as a group}, as reflected in the near zero average for two respective actions in Figures~\ref{figure5} and~\ref{figure6}, they do produce differences on individual turbines.  Some of the turbines are benefited from the technical actions, even P1. As a result certain subset of the turbines turns green after the first technical action.

\begin{figure}
		\subfigure[2015]{\includegraphics[width=0.5\textwidth]{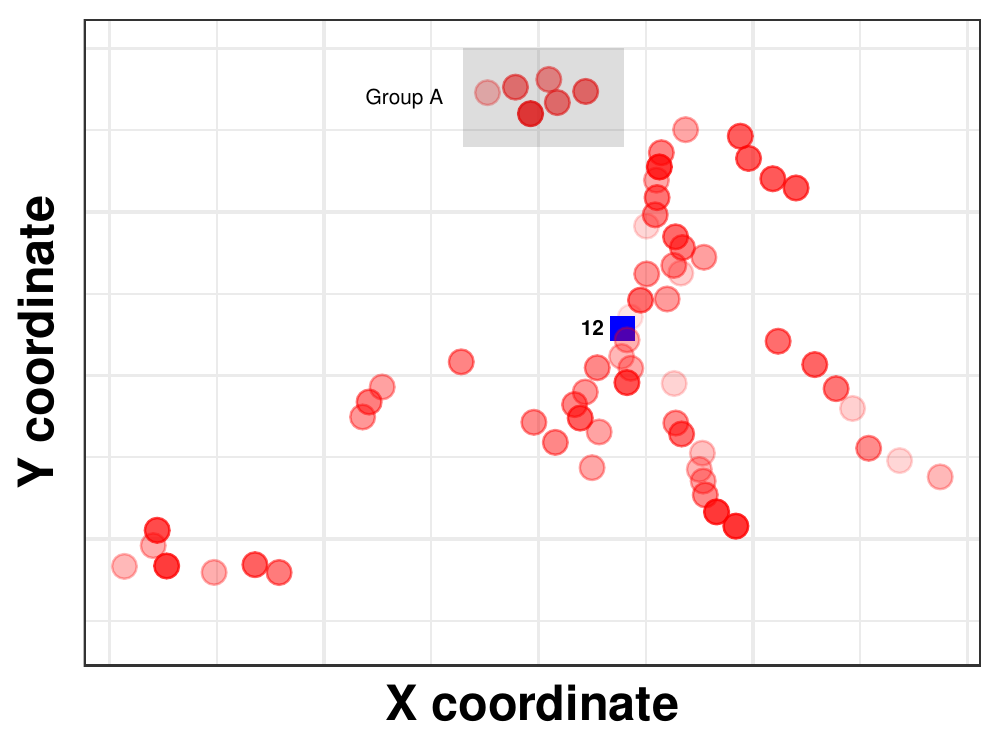}}
		\subfigure[2016]{\includegraphics[width=0.5\textwidth]{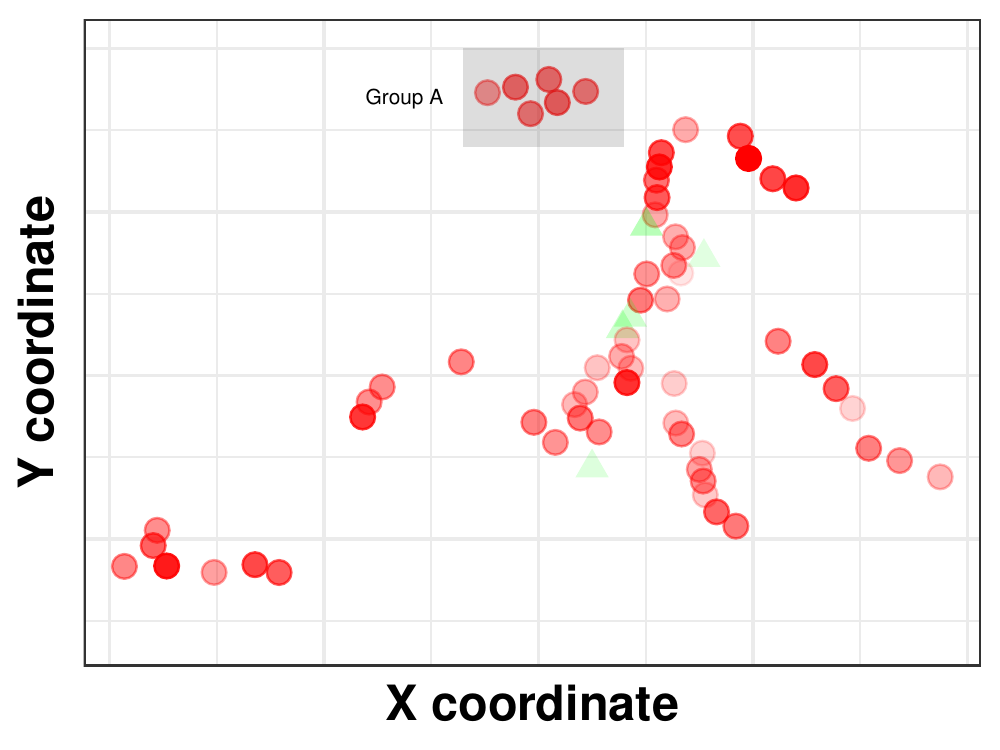}}
		\subfigure[2017]{\includegraphics[width=0.5\textwidth]{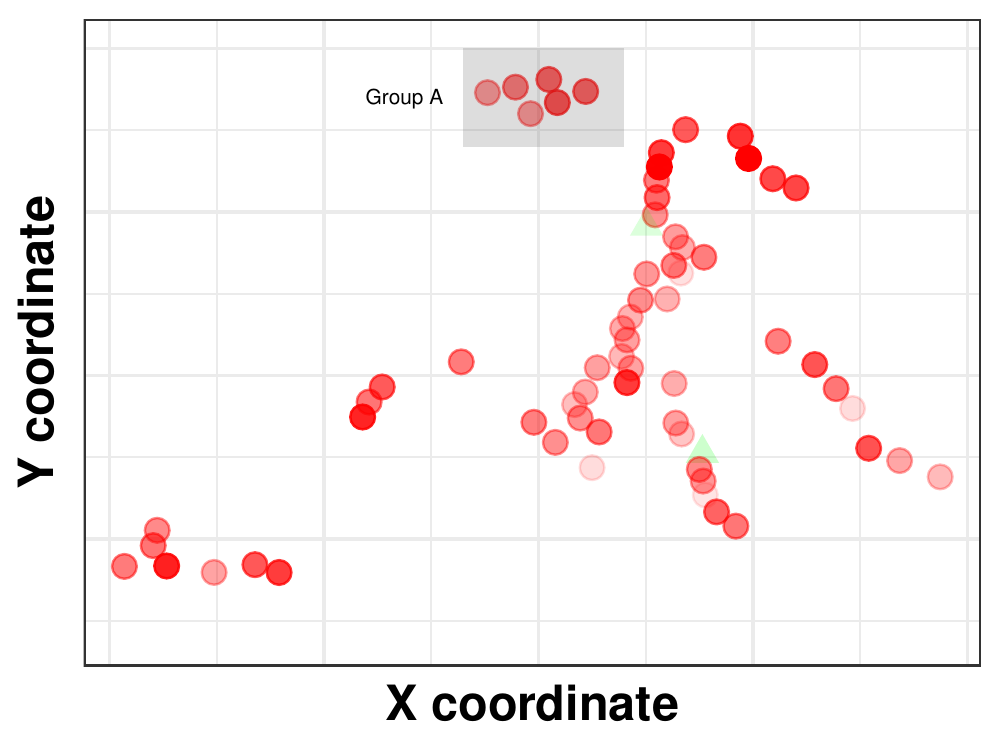}}
		\subfigure[2018]{\includegraphics[width=0.5\textwidth]{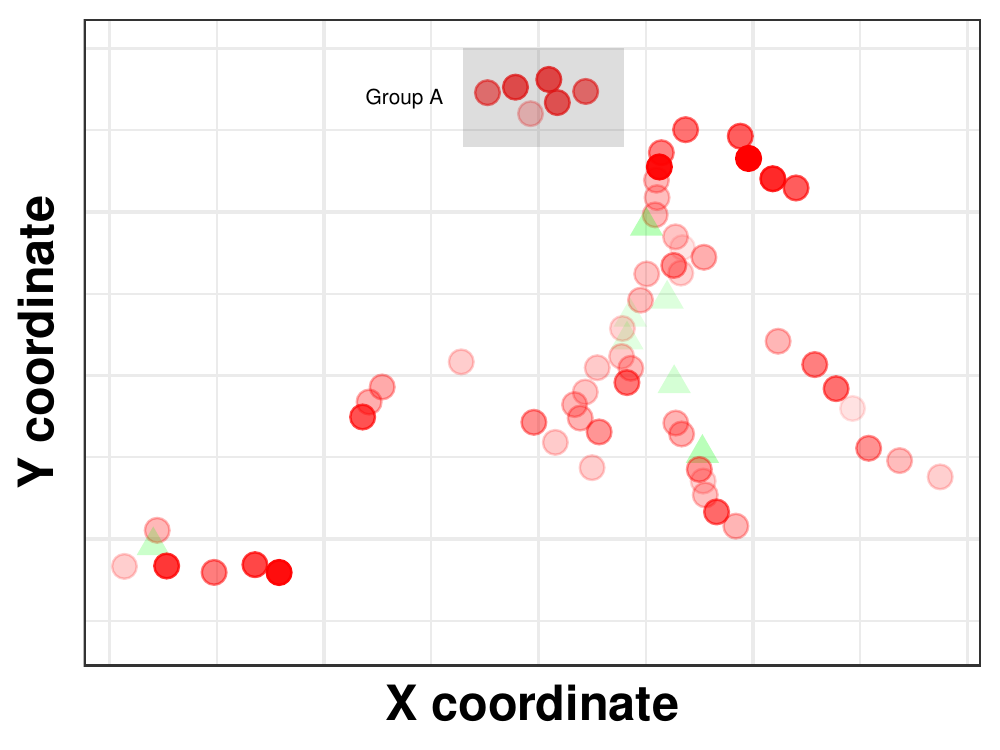}}
		\caption{Space-time performance comparison of wind turbines. Time period is each calendar year.}\label{figure8}
	\end{figure}

\begin{figure}
		\subfigure[Period 1]{\includegraphics[width=0.5\textwidth]{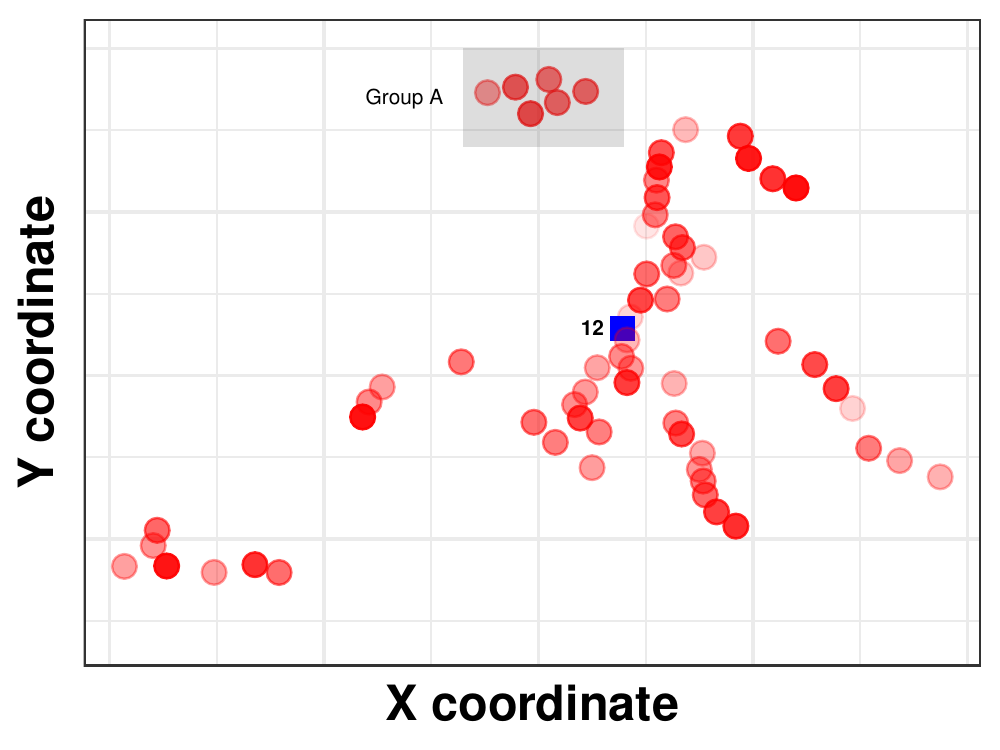}}
		\subfigure[Period 2]{\includegraphics[width=0.5\textwidth]{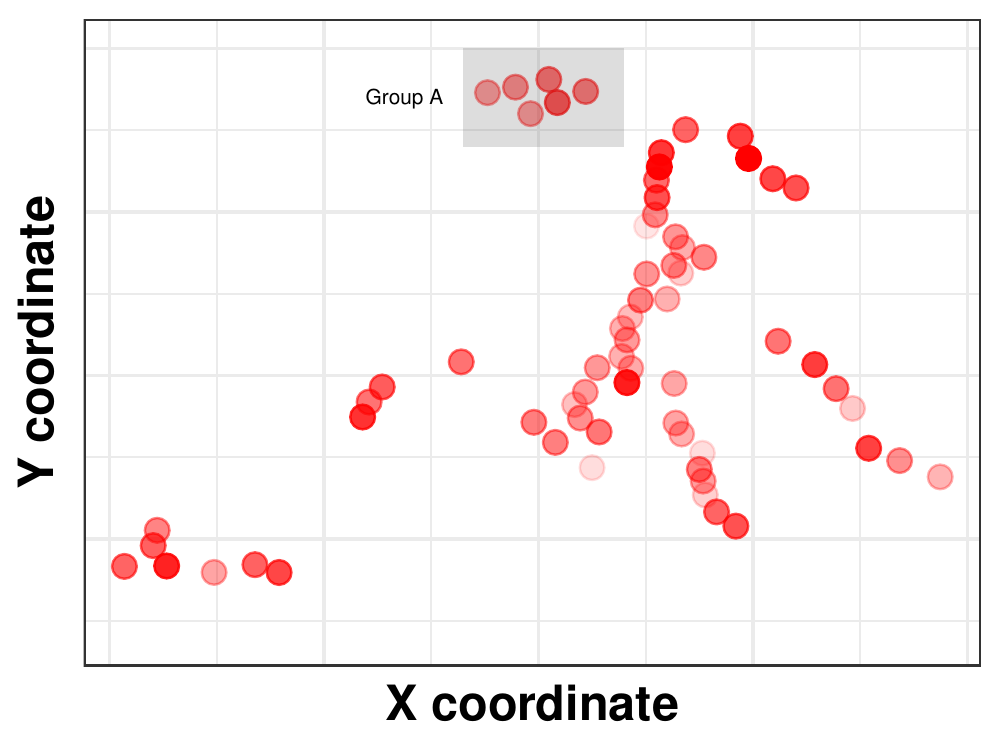}}
		\subfigure[Period 3]{\includegraphics[width=0.5\textwidth]{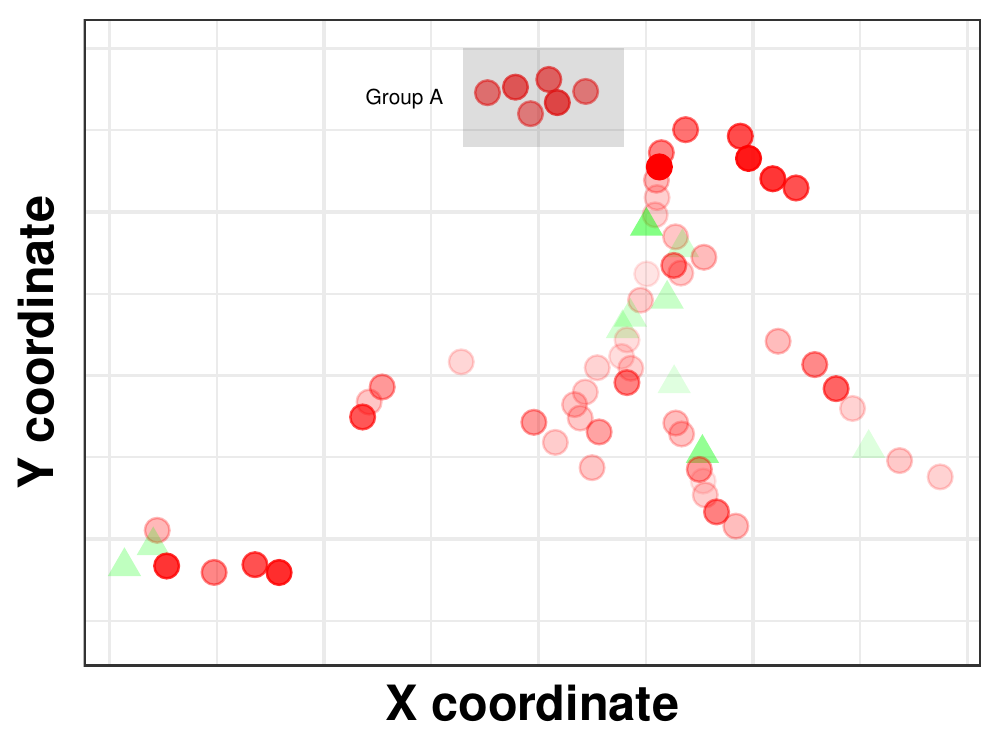}}
		\subfigure[Period 4]{\includegraphics[width=0.5\textwidth]{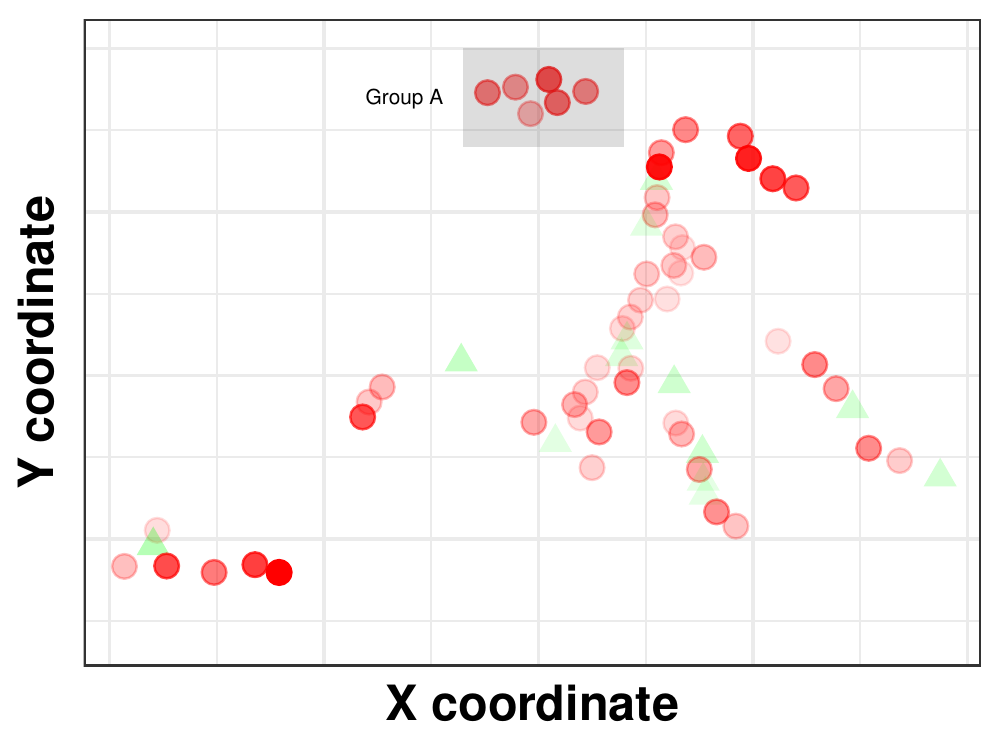}}
		\caption{Space-time performance comparison of wind turbines. Time period is those partitioned by the three technical actions.}\label{figure9}
	\end{figure}

	Figures~\ref{figure8} and~\ref{figure9} are the true space-time analysis, because the baseline is not reset every year. Rather, the baseline is the performance of \#12 of the first time period.  The difference between the two sets of figures is that Figure~\ref{figure8} uses a calendar year-based time period partition, so that the first time period is the year of 2015, whereas Figure~\ref{figure9} uses the time periods partitioned by the three technical actions explained in Figure~\ref{figure2}.
	
	In Figures~\ref{figure8}--\ref{figure9}, the set of turbines in Group A are in red for all periods, and some of the turbines are in darker red.  This initially comes as a surprise to the owner/operator's engineering team because based purely on power production, this set of turbines are deemed ``good'' as they produce more energy than most of other turbines. A closer look reveals that this group of turbines is on a hilltop, which is the highest elevation point on the wind farm. Understandably, their wind resources are better than other turbines at lower elevations.  The reason that Group A turbines produce more energy is not because they are more efficient but because they are better positioned.  Conditioned on the same wind and environmental conditions, this set of turbines is in fact worse off than many other turbines.  Had Group A turbines been operated as efficiently as \#12, they would have produced more energy due to the good wind resourses they enjoyed.
	
	\section{Summary}\label{section5}
	This paper presents a case study of performance analysis and comparison for all turbines on a moderate size wind farm for a period of four years. This study is rather different from a single turbine performance evaluation, which occupies most of the existing literature.  We believe that our space-time performance analysis is the first of its kind. The analysis results shed valuable insights to owners/operators, as they provide a comprehensive and quantitative picture of how things are going on the wind farm. Please note that the focus of this paper is to present a principled procedure for conducting this type of analysis. The question of how the analysis results should be taken to advise operations depends heavily on individual owner/operator's specific interests, objectives and needs and should be decided on a case-by-case basis.
	
	We want to note that under the advisement of the owner/operator, we use the nacelle data in our analysis.  We understand that there is some concern regarding the use of nacelle data. As the wind measurements on the nacelle are in the wake of the rotor, using them introduces error and uncertainty.  The alternative is to use the data from the met mast.  The reason that we did not use the mast data is because there is only a single mast on the farm.  When we compare the two cases, i.e., using the nacelle data as input versus using the mast data for constructing power curve models, we find that using the mast data, which may benefit the turbine closest to it, delivers a worse result for the whole-farm comparison.  The results are not even close---using the nacelle data typically gets the RMSE of power curve models down to 2--3\%, while using the mast data results in an 8--12\% RMSE.  The message is clear.  Unless one has a lot of met masts on a wind farm or has other means to measure free-stream wind for individual turbines (like using a LiDAR), the option of using nacelle data is, so far, practically the best.
	
	Our above argument is not to diminish the importance of the measurement accuracy.  In light of the error and uncertainty caused by the nacelle measurements, which could be up to several percent~\cite{IEC12}, we believe it is particularly important to use the statistical significant difference metric.  This is a consequence of the fact that the statistical significant difference metric is more robust and generalizable to other cases because it eliminates random fluctuation.  We hope that an economical solution for accurate wind measurements will become reality soon. Nevertheless, even when that happens, one still needs a principled data science procedure to conduct the space-time quantification and comparison for turbine performances.  In this sense, the proposed procedure in this paper could stay intact with only the input data replaced by more accurate ones.
	
	\section{Acknowledgement}
	Texas A\&M University's research team is partially supported by NSF grant IIS-1741173 and Goldwind China under contract no. M1900794.

	
	\bibliography{bibtex_dswe}
	
\end{document}